\documentclass[conference,a4paper,10pt]{IEEEtran}
\usepackage[pdftex]{graphicx}
\usepackage{amsmath,amssymb}
\usepackage{flushend}

\def\(({\left(}
\def\)){\right)}
\def\[[{\left[}
\def\]]{\right]}

\newcommand{\bx}{{\textbf {x}}}

\newcommand{\bh}{{\textbf {h}}}

\newcommand{\bre}{{\textbf {e}}}
\newcommand{\by}{{\textbf {y}}}
\newcommand{\bytilde}{{\tilde{\textbf {y}}}}

\newcommand{\be}{\begin{equation}}
\newcommand{\ee}{\end{equation}}
\newcommand{\bea}{\begin{eqnarray}}
\newcommand{\eea}{\end{eqnarray}}

\begin{document}

\sloppy

\title{Robust error correction for real-valued signals via message-passing
  decoding and spatial coupling}

\author{
\IEEEauthorblockN{Jean Barbier and Florent Krzakala$^*$}
\IEEEauthorblockA{
ESPCI and CNRS UMR 7083 \\
10 rue Vauquelin,\\
Paris 75005  France\\
$*$ IEEE member, fk@espci.fr}
\and
\IEEEauthorblockN{Lenka   Zdeborov\'a}
\IEEEauthorblockA{Institut de Physique Th\'eorique\\ IPhT, CEA Saclay\\ and URA 2306,
CNRS\\ 91191 Gif-sur-Yvette, France.}
\and
\IEEEauthorblockN{Pan Zhang}
\IEEEauthorblockA{
ESPCI and CNRS UMR 7083 \\
10 rue Vauquelin,\\
Paris 75005  France}
}


\maketitle

\begin{abstract}
  We revisit the error correction scheme of real-valued signals when
  the codeword is corrupted by gross errors on a fraction of entries 
  and a small noise on all the entries. 
  Combining the recent developments of approximate message passing
  and the spatially-coupled measurement matrix in compressed sensing
  we show that the error correction and its robustness towards noise
  can be enhanced considerably. We discuss the performance in the
  large signal limit using previous results on state evolution,  
  as well as for finite size signals through numerical
  simulations. Even for relatively small sizes, the approach proposed
  here outperforms convex-relaxation-based decoders. 
\end{abstract}

\section{Introduction}
Although information is discrete in the classical coding theory, there are
situations of interest where one should consider real-valued signals,
such as scrambling of discrete time analog signals for privacy
\cite{1056050}, network \cite{feizi2011power,shintre2008real} or
jointed source and channel coding \cite{grangetto2005joint}, or in the
impulse noise cancellation in orthogonal frequency division
multiplexing system \cite{4595196}. To perform error correction for
such real signals over a channel with a gross errors on a fraction of
elements (and small noise for all of them) a compressed-sensing-based
scheme has been proposed by Donoho and Huo~\cite{959265} and Candes
and Tao~\cite{CandesTao:05}. Here we reconsider this problem, taking
full advantage of the recent progresses in compressed sensing theory.

The problem is easily stated. One is given a real-valued signal $\bx$,
and a channel that adds gross errors to a fraction of elements. Is
there a way to encode the signal such that the errors added by the
channel can be corrected? Can this approach still be used when the
channel is in addition adding a small noise to all elements (a
situation arguably much closer to some real
channels~\cite{959265,Candes:2008:HRE:2263476.2273354,lampe2011bursty})?
The method proposed in
\cite{959265,CandesTao:05,Candes:2008:HRE:2263476.2273354} is to first
multiply the signal $\bx$ by a random matrix in order to create a
codeword of larger dimension, and then to use the classical compressed
sensing approach, based on convex-relaxation decoding, to correct the
errors of transmission.

Our contribution in the present paper is three-fold. (1) We replace
the convex-relaxation decoding by the Bayesian Approximate
Message-Passing (AMP) decoder that uses the available prior
information about the error vector
\cite{DonohoMaleki10,KrzakalaPRX2012}. This provides a significant
improvement in performances. (2) We consider a quasi-sparse channel
where, in addition to the gross errors on a fraction of elements, there
is a small additive random white noise, and in the lines of
\cite{BarbierKrzakalaAllerton2012} show that the performance of AMP
decoder is stable under this additional noise.  (3) We use spatially-coupled
measurement matrices in the decoding
\cite{KrzakalaPRX2012,DonohoJavanmard11}, which allows to further
enhance the possibility for error correction (and up to its
information-theoretical limit in the case of strictly sparse
noise). 

Our paper relies on the development of the Bayesian AMP
algorithm
\cite{DonohoMaleki10,KrzakalaPRX2012,Rangan10b,KrzakalaMezard12},
whose behavior for large signal sizes can be studied rigorously using
the state evolution technique
\cite{BayatiMontanari10,bayati2012universality}; on the development of
spatially coupled error correcting codes on binary variables, see
\cite{KudekarRichardson12} for a review, and related measurement
matrices in compressed sensing
\cite{KrzakalaPRX2012,DonohoJavanmard11,KudekarPfister10}; and on the
use of quasi-sparsity in the Bayesian approach
\cite{BarbierKrzakalaAllerton2012}. While the analysis of the
reconstruction performance of AMP are rigorous in the large signal
limit, we also consider the performance for finite size signals
through numerical experiments.

\section{Compressed-sensing Based Error-Correction}
Consider a real-valued vector of information $\bx \!\in\!
\mathbb{R}^N$, encode this vector by a full-rank real $M\! \times \!N$
matrix $A$, with $R\!=\!N/M\!<\!1$ being the coding rate (the
redundancy rate introduced in the code), so that the encoded vector is
$\by=A\bx \in \mathbb{R}^M$. Since $A$ is full rank, one can recover
the original signal $\bx$ from the encoded one multiplying by the
pseudo-inverse. The encoded signal is sent through a noisy channel
and gives rise to the corrupted codeword $\bytilde=\by+\bre$ where the
element of $\bre$ are iid with a distribution \be P(e_i) = \rho {\cal
  N}(0,1+\epsilon)+ (1-\rho) {\cal N}(0,\epsilon) \, ,  \label{noise}
\ee where $0<\rho<1$. We thus have a fraction $\rho$ of elements with
gross (variance $1+\epsilon$) errors, the rest having small (variance
$\epsilon$) amplitudes. One then considers a full rank
"parity-check"-like matrix $F$ such that $F A=0$. We construct such a
pair of matrices by first choosing a $P \times M$ matrix $F$ with
independent normally distributed elements of zero mean and variance
$1/M$ (or variance specified by the seeding matrix, see later), the
kernel of $F$ is then the range of the encoding matrix $A$
\footnote{Note that
  \cite{CandesTao:05,Candes:2008:HRE:2263476.2273354} take $A$ as the
  random matrix, we choose the opposite in order to be able to implement the
  spatially coupled decoding.}. One must have $P \le M-N$, in order to
maximize the coding rate $R=N/M$, we take from now on $P=M-N$.
The application of $F$ to the corrupted signal $\bytilde$ results in
the real-valued vector
\be \bh= F (\by+\bre) = F \bre \, ,
\label{CS-eq}
\ee
where $\bh$ has dimension $M-N$ and $\bre$ is a quasi-sparse vector of
dimension $M$. This is a compressed sensing problem: reconstruct the
quasi-sparse $M$-dimensional error $\bre$ given $M-N$ of its linear
projections (measurements) $\bh$. In the context of compressed sensing $F$
is the measurement matrix. 

\begin{figure}
\centering
\includegraphics[width=0.5\textwidth]{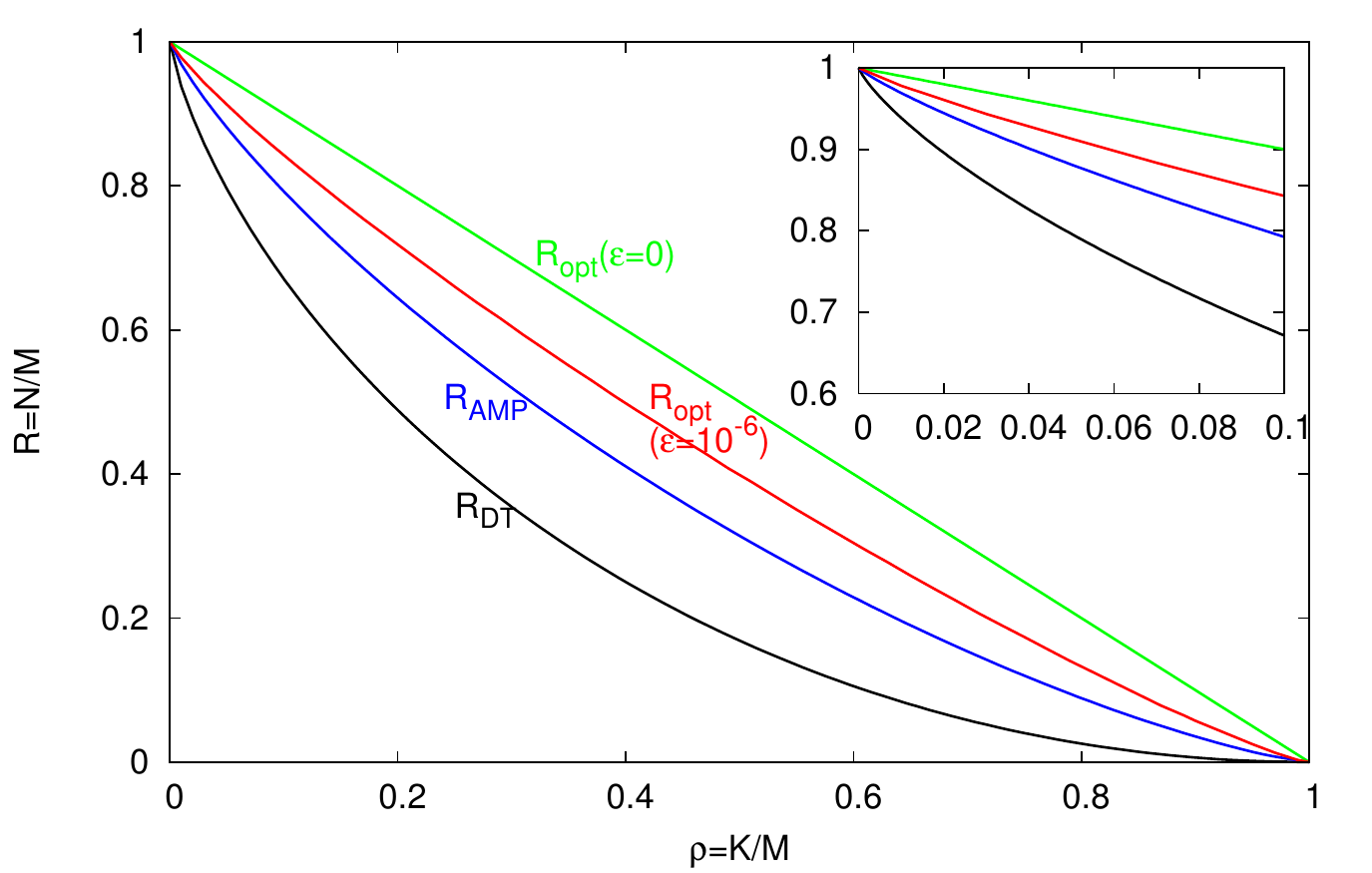}
\caption{\label{fig:PhaseDiagram} Phase diagram showing the coding
  rate $R=N/M$ below which error correction can be performed over a channel
  with noise described by eq. (\ref{noise}), plotted as a function of
  the noise sparsity $\rho$, zoom in the inset. The black (bottom)
  curve $R_{\rm DT}$ depicts the limit of performance of the
  $\ell_1$-minimization approach for $\epsilon=0$. The blue (2nd from
  bottom) curve shows the limit of performance of the Bayesian AMP
  approach for $\epsilon=0$, note that up to about $\epsilon\lesssim
  10^{-5}$ this curve does not change visibly. The green (top)
  curve, given by $R_{\rm opt}=1-\rho$, depicts the highest
  possible coding rate for which exact decoding is possible for
  $\epsilon=0$. Below the red (2nd from top) line error correction
  with MSE comparable to $\epsilon=10^{-6}$ is possible with the Bayes
  optimal estimation of the error vector. These two rates $R_{\rm
    opt}$ can be reached in the limit of large signal size using the
  seeded Bayesian AMP approach.}
\end{figure}

Let us first review the possibility of this error-correction scheme
when the error $\bre$ is exactly sparse, i.e. $\epsilon=0$.  Using an
intractable $\ell_0$ minimization, the gross error $\bre$ in
eq.~(\ref{CS-eq}) can be found exactly as long as $M-N>M\rho$. So
error correction in real valued signals corrupted by sparse gross
noise is possible (but may be hard) for coding rates $R<R_{\rm
  opt}=1-\rho$. Popular tractable $\ell_1$ minimization, as used in
\cite{CandesTao:05,Candes:2008:HRE:2263476.2273354}, recovers the
error $\bre$ exactly when $M-N \ge \alpha_{\rm DT} M$, where
$\alpha_{\rm DT} $ is the Donoho-Tanner measurement rate
\cite{Donoho05072005}. This means that the coding rate must be lower
than $R \le R_{\rm DT} = 1-\alpha_{\rm DT}$. These two transitions are
depicted in Fig.~\ref{fig:PhaseDiagram} and one can see that $R_{\rm
  DT}$ is considerably lower than $R_{\rm opt}$. A first step to
improvement is to decode with an approximate message passing approach.

\section{Decoding with approximate message passing}
Consider the compressed sensing problem in eq.~(\ref{CS-eq}) to
reconstruct the vector $\bre$ based on the knowledge of the
$M-N$-dimensional $\bh$ and the matrix $F$. When the distribution of
the error elements is known, the Bayes optimal way of
estimating $\hat \bre$ that minimizes the MSE
$E=\sum_{i=1}^M (e_i - \hat e_i)^2 /M$ with the true error $\bre$
is given as \be \hat e_i=\int \text{d} e_i \, e_i\, \nu_i(e_i) \,
,\label{average_marginal} \ee where $\nu_i(e_i)$ is the marginal
probability distribution of the variable $i$
\be
\nu_i(e_i) \equiv \int_{\{e_j\}_{j\neq i}} P(\bre| \bh)  \prod_{j \neq
  i} de_j\ee
under the posterior measure \be P(\bre | \bh) = \frac{1}{Z(\bh)} P(\bre)
\prod_{\mu=1}^{M -N}\delta(h_\mu - \sum_{i=1}^M F_{\mu i}e_i)\,
.\label{p_bayes} \ee 
Such an optimal Bayes reconstruction is not computationally tractable in general and
in order to get an estimate of the marginals $\nu_i(e_i)$, we use the
AMP algorithm.

\subsection{AMP reminder}
AMP is defined over a graphical model and iteratively updates
pre-estimates of the mean and variance $R_i$, $\Sigma_i^2$ and {\it a
  posteriori} estimates of the mean and variance $a_i$, $v_i$ for each
component $i$ of $\hat \bre$. They are updated as follows (for the
derivation in the present notation see \cite{KrzakalaMezard12} and for
the original one \cite{Rangan10b,DonohoMaleki10}) where, for
convenience, we have also defined for every component $\mu$ of $\bh$
auxiliary variables $V_\mu$ and $\omega_\mu$:
\bea
V^{t+1}_\mu &=& \sum_i F_{\mu i}^{2} \, v^{t}_i \, , \label{TAP_ga} \\
\omega^{t+1}_\mu &=& \sum_i F_{\mu i} \, a^t_i -\frac{
  (h_\mu-\omega^t_\mu)}{V^t_\mu} \sum_i F_{\mu i}^2\,
v^t_i \, , \label{TAP_al}  \\
(\Sigma^{t+1}_i)^2&=&\left[ \sum_\mu \frac{F^2_{\mu i}}{ V^{t+1}_\mu}
\right]^{-1}\, ,
      \label{TAP_U}\\
      R^{t+1}_i&=& a^t_i + \frac{\sum_\mu F_{\mu i} \frac{(h_\mu - \omega^{t+1}_\mu)}{ V^{t+1}_\mu}}{ \sum_\mu \frac{ F_{\mu i}^2}{V^{t+1}_\mu}}\, , \label{TAP_V}\\
      a^{t+1}_i\!\! &=&\!\! f_a\left((\Sigma^{t+1}_i)^2,R^{t+1}_i\right) ,    v^{t+1}_i\!\!=\!\!f_c\left((\Sigma^{t+1}_i)^2,R^{t+1}_i\right) \, .\nonumber
\eea
Where, using $P(\bre)=\prod_i P(e_i)$ as given in eq.~(\ref{noise}), the functions $f_a$ and $f_c$ read \cite{BarbierKrzakalaAllerton2012}
\begin{align}
&f_a(\Sigma^2,R) = \frac{  \sum_{a=1}^2 w_a
e^{-\frac{R^2}{2(\Sigma^2+\sigma_a^2)}}
\frac{ R \sigma_a^2}{(\Sigma^2+\sigma_a^2)^{\frac{3}{2}}} }{   \sum_{a=1}^2 w_a \frac{1}{\sqrt{\Sigma^2+\sigma_a^2}}
e^{-\frac{R^2}{2(\Sigma^2+\sigma_a^2)}} } \,
,\\
&f_b(\Sigma^2,R) = \frac{  \sum_{a=1}^2 w_a
e^{-\frac{R^2}{2(\Sigma^2+\sigma_a^2)}}
\frac{\sigma_a^2
  \Sigma^2 (\Sigma^2 + \sigma_a^2)+  R^2 \sigma_a^4 }{(\Sigma^2+\sigma_a^2)^{\frac{5}{2}}} }{    \sum_{a=1}^2 w_a \frac{1}{\sqrt{\Sigma^2+\sigma_a^2}}
e^{-\frac{R^2}{2(\Sigma^2+\sigma_a^2)}} } \nonumber \,
, \\
&f_c(\Sigma^2,R)=f_b(\Sigma^2,R)-f_a^2(\Sigma^2,R) \, .
\end{align}
with
\bea
w_1=\rho\, , \quad
\sigma_1^2=1 +\epsilon \, , 
w_2=1-\rho\, ,\quad \sigma_2^2=\epsilon\, .  
\eea

The initialization is set as: $a_i^{t=0}=0$,
$v_i^{t=0}=\rho+\epsilon $,
$\omega_\mu^{t=0}=h_\mu$.
Once the convergence of the iterative procedure is obtained the
estimate of the $i$th component of the error $\hat \bre$ is $a^{t}_i$. 
\subsection{Performance of AMP}
As shown by Bayati and Montanari \cite{BayatiMontanari10}, in the limit of large system sizes,
i.e. when parameters $\rho, \epsilon,R$ are fixed whereas $M\to
\infty$, the evolution of the AMP algorithm can be described exactly
using the ``state evolution''. This allows to
evaluate the MSE achieved by the AMP reconstruction.  When $F$ is an
homogeneous random iid matrix the state evolution is written in terms
of MSE at iteration-time $t$, which evolves as (for a derivation
see e.g. \cite{Rangan10b,KrzakalaMezard12,BayatiMontanari10})
\be
    E^{t+1}\! =\!\! \sum_{a=1}^{2} w_a  \int {\cal D}z f_c\left( \frac{1}{\hat
      m^t}, z\sqrt{\sigma_a^2  + \frac{1}{\hat m^t} }\right)\!\!,
  {\hat m}^t = \frac{\alpha}{E^{t}}\, ,  \label{Et}
\ee
where ${\cal D} z = dz  e^{-z^2/2}/\sqrt{2\pi} $ is a Gaussian measure
for the integral, and where $E^{t=0}=\rho+\epsilon $. 

The analysis of the Bayesian AMP for Gauss-Bernoulli noise $\bre$ (that is,
with $\epsilon=0$) has been considered in great details in
\cite{KrzakalaPRX2012,KrzakalaMezard12}. In that case AMP reconstructs
{\it perfectly} the solution in a region larger than the
$\ell_1$-minimization and up to the so-called spinodal transition
$\alpha_{\rm AMP}$. In the notation of the present problem, where this
leads exact decoding for considerably larger coding rates: The
resulting $R_{\rm AMP}=1-\alpha_{\rm AMP}$ is shown in blue in
Fig.~\ref{fig:PhaseDiagram} (with data adapted from
\cite{KrzakalaPRX2012}) where the advantage over the
$\ell_1$-minimization decoding is clear. For a fraction of $\rho\!=\!0.1$
gross elements, for instance, the improvement goes from a necessary
coding rate $R \approx 0.67$ for $\ell_1$ to $R \approx 0.79$ for AMP.
It should be noted, however, that the $\ell_1$
performance is independent of the distribution of the gross error,
whereas the Bayes AMP uses the distribution of the elements of the
gross error. The properties of the channel are, however, often well
known, in which case the improvement depicted if
Fig.~\ref{fig:PhaseDiagram} is indeed achievable.

To assess how robust are these results towards approximately sparse
noisy channels (nonzero value of $\epsilon$ in eq.~(\ref{noise})) we
use the analysis that was performed in
\cite{BarbierKrzakalaAllerton2012}. It was shown that for about
$\epsilon\lesssim 10^{-5}$ and $\alpha>\alpha_{\rm AMP}$ the AMP
algorithm leads to reconstruction with MSE comparable to
$\epsilon$. This shows that the AMP approach is actually very robust
to such noise.

\section{Spatially coupled measurements}
Despite the advantage of the AMP-based decoding over the $\ell_1$-minimization, it is
still not asymptotically optimal since $R_{\rm AMP}< R_{\rm opt}$, and one ideally aims
to perform error correction with highest possible coding rates. In order to do so, we
shall mimic the strategy used in LDPC error correcting codes over
binary signals (see \cite{KudekarRichardson12} for a review) 
in compressed sensing based-codes over real-valued signals. It has been shown recently
\cite{KrzakalaPRX2012,DonohoJavanmard11} (first heuristically and
numerically, and later rigorously) that for compressed sensing
of sparse signals with known empirical distribution of components the
theoretically optimal reconstruction can be achieved in the large
system size limit with the combined
use of AMP algorithm \cite{DonohoMaleki10,Rangan10b} and seeding
(spatially coupled) measurement matrices
\cite{KrzakalaPRX2012,KudekarPfister10}. 

\begin{figure}[t]
\center\includegraphics[width=2.4in]{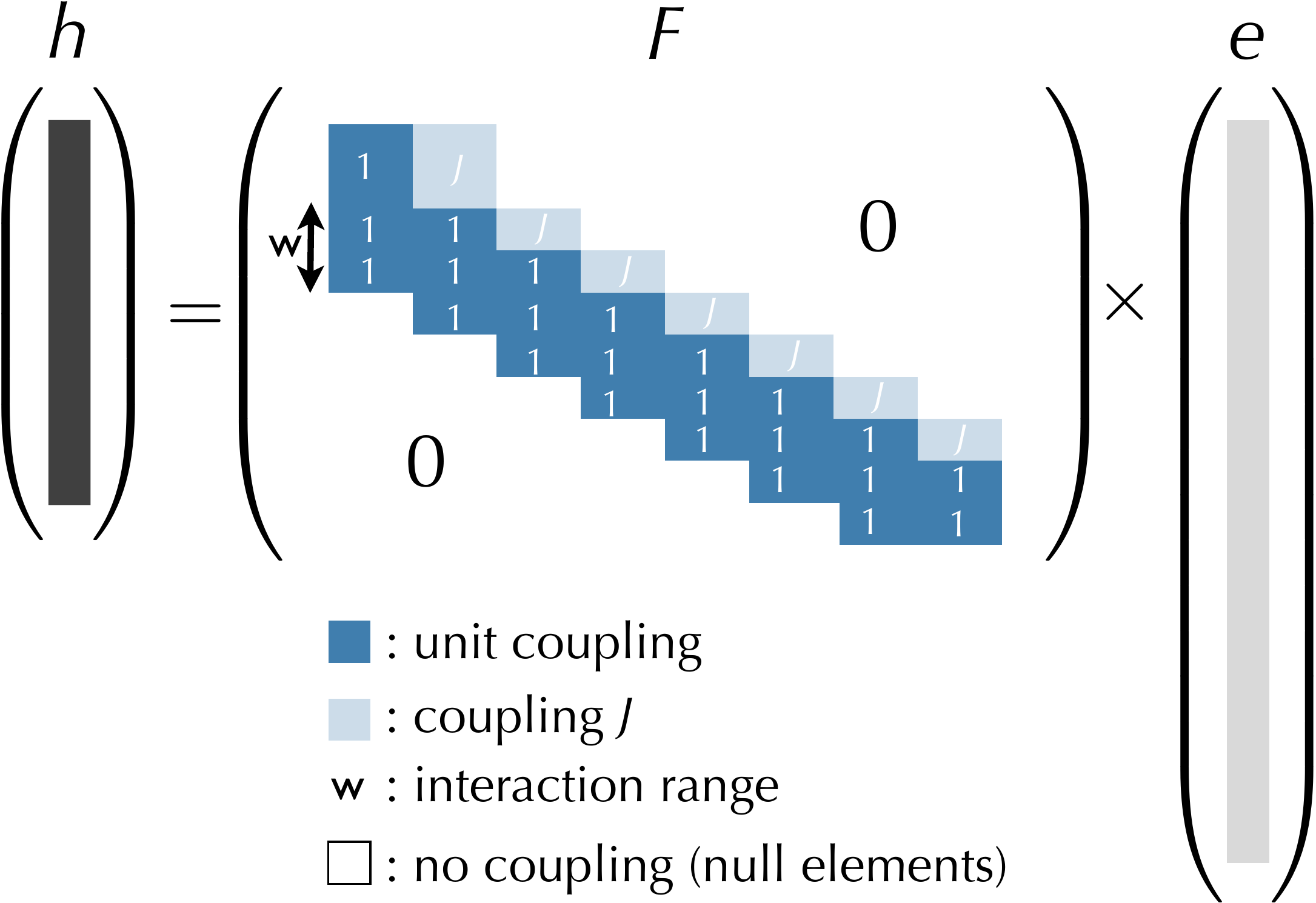}
\caption{Sketch of the seeding matrix $F$ we used to approach optimal
  reconstruction of the error vector with the AMP algorithm (see
  \cite{BarbierKrzakalaAllerton2012}). We use a number of
  variable-blocks $L_c=L$, and $L_r=L+1$ measurement blocks. The
  matrix components are iid with zero mean and variance $1/M$ for the
  blocks on the diagonal and for a number $W$ of lower diagonals, the
  upper diagonal blocks have components with variance~$J/M$.}
\label{fig_matrix}
\end{figure}

In the present framework of error correction of real-valued signals,
the spatial coupling can be implemented by first constructing the
matrix $F$ as the seeding matrix of \cite{KrzakalaPRX2012}, then 
determining the coding matrix $A$ as the null space of the matrix
$F$. We refer to
\cite{KrzakalaPRX2012,BarbierKrzakalaAllerton2012,DonohoJavanmard11,6483301}
for a general discussion of spatial coupling in compressed sensing.
The seeding measurement matrices $F_{\mu i} \in \mathbb{R}^{(M-N)\times N}$ that we
use in the rest of this paper are constructed as in
\cite{BarbierKrzakalaAllerton2012}, see Fig.~\ref{fig_matrix} for an example. The $M$
components of the error vector are sliced into $L_c$ equally-sized
groups and the $M-N$ measurements into $L_r$ groups, the first of
which having a larger number of measurements $m_{\rm seed}$ than the
others $m_{\rm bulk}$. Define $\alpha_{\rm seed}=L_c m_{\rm seed}/M$
and $\alpha_{\rm bulk}=L_c m_{\rm bulk}/M$. Define $\alpha=(M-N)/M$
as the total measurement rate, we have $\alpha= [\alpha_{\rm
    seed} +(L_r-1) \alpha_{\rm bulk}]/L_c $. The matrix $F$ is then
composed of $L_r\times L_c$ blocks where the elements $F_{\mu i}$ are
generated independently, such that if $\mu$ is in group $q$ and $i$ in
group $p$ then $F_{\mu i}$ is a random number with zero mean and
variance $J_{q,p}/M$. In this work we use seeding matrices illustrated
in Fig.~\ref{fig_matrix} parameterized by $L$, $W$, and $J$.

A rigorous analysis of the evolution of AMP reconstruction was
performed for such matrices, assuming the knowledge of the $L_r\times
L_c$ coupling matrix $J_{q,p}$ \cite{bayati2012universality}. The
state evolution for the block matrices can be done by defining
$E_p^{t}$ to be the mean-squared error in block $p$ at time $t$. Then,
$E_p^{t+1}$ depends on $\hat m_p^t$ from the same block according to
eq.~(\ref{Et}). The quantity $\hat m_p^t$ depends on the MSE $E^{t}_q$
from all the blocks $q=1,\dots,L_c$ as \be \hat m_p^t =
\frac{\alpha_{\rm seed}J_{1p}}{\sum_{q=1}^{L_c} J_{1q} E_q^t } +
\alpha_{\rm bulk}\sum_{r=2}^{L_r} \frac{J_{rp}}{\sum_{q=1}^{L_c}
  J_{rq} E_q^t }\, . \ee This kind of state evolution belongs to the class
for which threshold saturation (asymptotic achievement of performance
matching the optimal Bayes estimation of the error) was proven in
\cite{YedlaJian12} (when $L\to \infty$, $W\to \infty$ and $L/W \gg
1$).  It is also possible to compute the possible performance of
Bayes-optimal reconstruction of the error analytically in the $N \to \infty$ limit
as done in \cite{BarbierKrzakalaAllerton2012}, and the results are
shown again in Fig.~\ref{fig:PhaseDiagram} in green (upper-most) curve
for $\epsilon=0$ and in red (2nd from top) curve for
$\epsilon=10^{-6}$. The conclusion is that with the seeding matrices
$F$, one can perform error correction using AMP up to these high coding
rates.

\begin{figure}[!ht]
\centering
\includegraphics[width=0.5\textwidth]{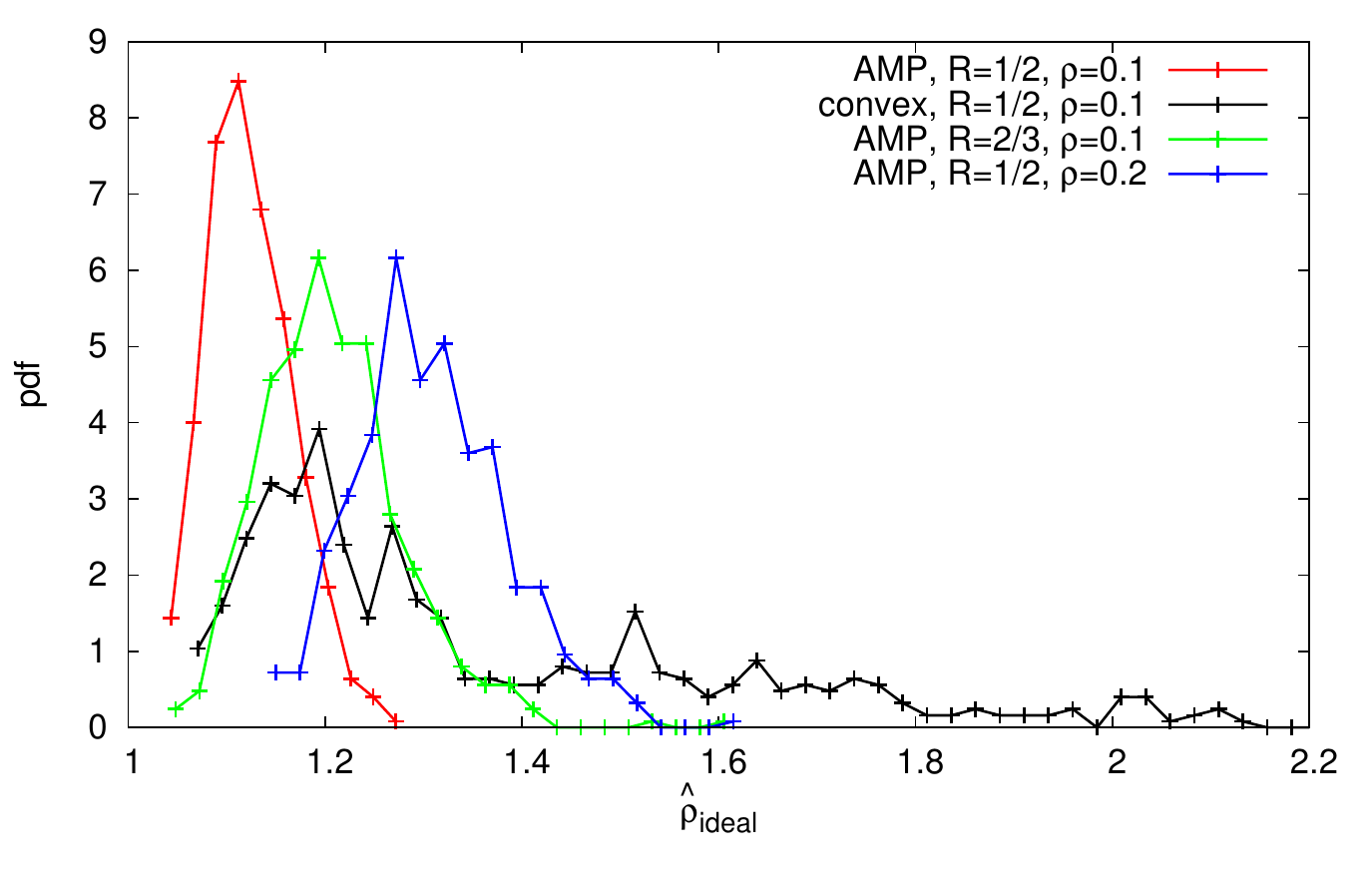}
\caption{\label{fig:dist:rho} Robustness to noise of the error
  correction for signal of size
  $N=256$. We compare the performances of the AMP-based and $\ell_1$ decoders,
  using $\epsilon=10^{-6}$ and homogeneous matrices $F$. The figure
  shows the probability density function (estimated with $500$ instances) of the
  robustness ratio eq.~(\ref{rhoideal}), called ${\hat \rho}_{\rm ideal}$ in
  \cite{Candes:2008:HRE:2263476.2273354} at different values of
  the coding rate $R$ and gross noise sparsity $\rho$. For
  coding rate $R=1/2$ and gross noise sparsity
  $\rho=0.1$, both methods (AMP in red and $\ell_1$ in black)
  are giving values close to one. However, AMP is better, on
  average it gives $0.89$ versus $0.74$ for $\ell_1$. Furthermore, AMP still
  performs very well when the fraction of gross errors is double
  (blue curve, with $\rho=0.2$) or when the coding rate~$R$ is higher
  (green curve, with $R=2/3$). In both these cases, the
  $\ell_1$-based reconstruction gives poor results, an
  average value  ${\hat \rho}_{\rm ideal} =37.1$ for
  $\rho=0.2$, $R=1/2$, versus $1.30$ for AMP; and ${\hat \rho}_{\rm ideal} = 18.3$ for
  $\rho=0.1$, $R=2/3$, versus $1.20$ for AMP.}
\end{figure}

\section{Numerical tests}

The asymptotic guarantees given in the last sections are encouraging,
but evaluating analytically finite $M$ correction is intrinsically
more difficult and hence we withdraw to numerical verifications of the
achievable coding rates for sizes relevant for practical
applications. Our Matlab code and demo are available on
http://aspics.krzakala.org.

For numerical verifications, we used $N$-dimensional Gaussian signal
$\bx$ with zero mean and unit variance (the algorithm is {\it not}
using this information) and a channel noise distributed according to
eq.~(\ref{noise}). We performed the Bayesian AMP algorithm to
estimate the error $\hat \bre$. For exactly sparse channel, $\epsilon=0$,
the exact reconstruction of $\bre$ is possible and hence $\bx$
can be recovered exactly. For approximately sparse channel,
$\epsilon>0$, we use the AMP estimate of the error $\hat \bre$ to compute
the estimate of $A \hat \bx$ and finally a pseudoinverse of $A$ to estimate 
the signal $\hat \bx$. We compare to the $\ell_1$ decoding approach
(including the reprojection step) as developed in
\cite{CandesTao:05,Candes:2008:HRE:2263476.2273354}.

\begin{figure}[t]
\centering
\includegraphics[width=0.5\textwidth]{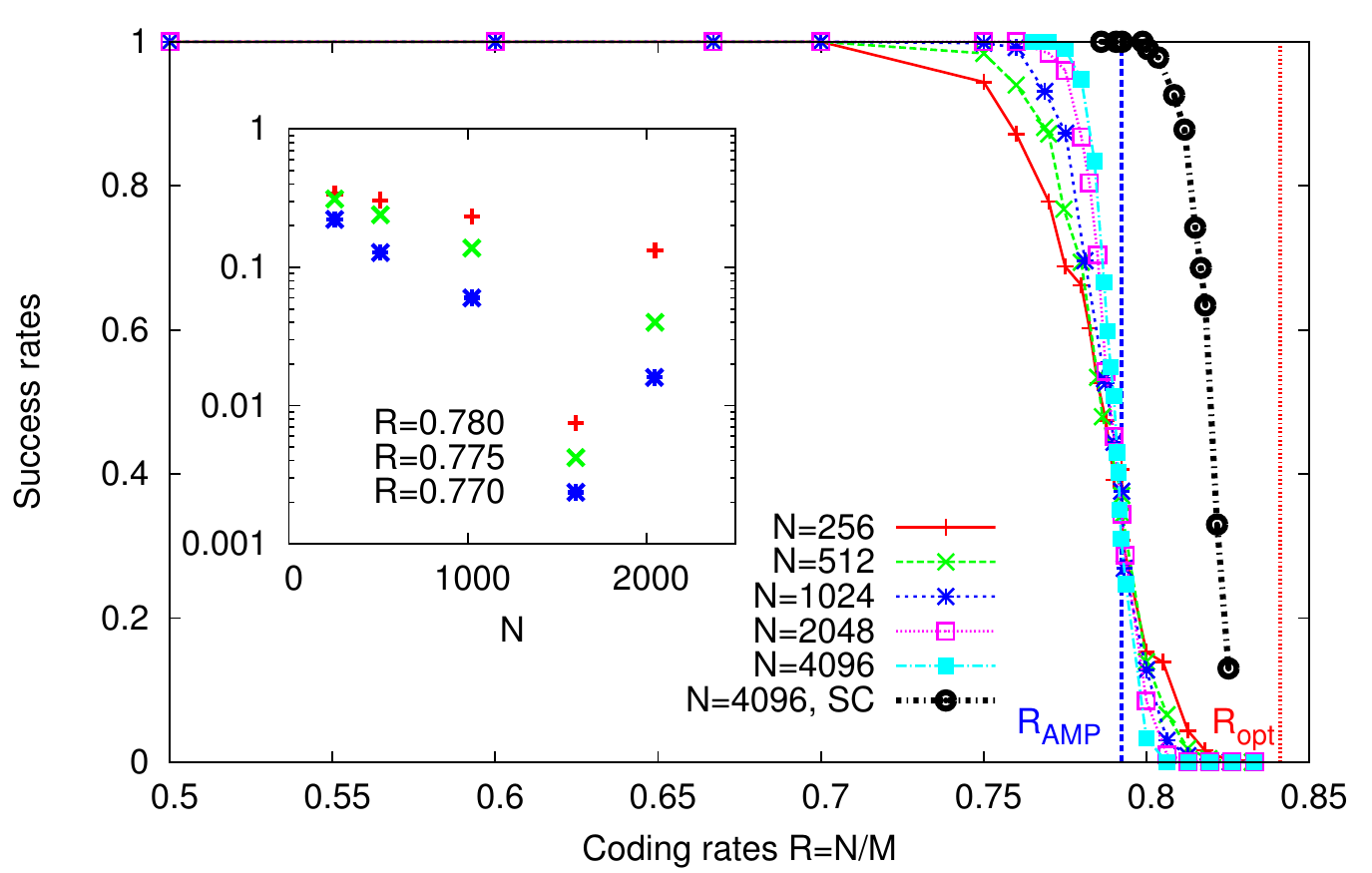}
\caption{\label{fig:frac} Success rates of the decoding over $500$
  instances for different signal sizes $N$ with noise parameters
  $\rho=0.1$, $\epsilon=10^{-6}$, both with seeding matrices $F$ (SC)
  and homogeneous random ones, as a function of the coding rate
  $R$. The vertical lines represent the limiting asymptotic
  coding rate for AMP with homogeneous and seeding matrices
  respectively. The maximum number of iterations in these simulations
  is set to $1000$. An instance is considered as successful if the final
  mean-squared-error of the reconstructed signal is less than
  $10^{-5}$. The inner plot shows the probability of failure
  that decays exponentially with the signal size for three different
  values of coding rate~$R$.}
\end{figure}

The data for $N\!=\!256$ are shown in Fig.~\ref{fig:dist:rho} where the
performance of our Bayes AMP decoding algorithm is compared to the
$\ell_1$-based decoding of \cite{Candes:2008:HRE:2263476.2273354}.
Following \cite{Candes:2008:HRE:2263476.2273354} we introduce an
estimator of the robustness to noise called ${\hat \rho}_{\rm ideal}$ as 
the ratio of the MSE of the
reconstructed signal $\hat \bx$ with the MSE of the "ideal"
reconstruction $\hat \bx_{\rm ideal}$,
where the pseudoinverse of $A$ is applied to $\by$ that was corrupted
only by the small additive noise without gross errors
\be
{\hat \rho}_{\rm ideal} = \frac{||\hat \bx -  \bx||_2}{||\hat \bx_{\rm ideal} -
  \bx||_2}\, .
\label{rhoideal}
\ee 
Fig.~\ref{fig:dist:rho} depicts the histogram of ${\hat \rho}_{\rm
  ideal}$ over 500 random instances of the problem. 
We find that in all the cases we have tried with AMP (which were all in
the favorable region of the asymptotic phase diagram), the robustness
estimator ${\hat \rho}_{\rm ideal}$ is very close to unity, even at
these relatively small sizes.  Moreover the robustness estimator of
AMP was always on average closer to unity than the one based on
$\ell_1$ estimation and the distribution more peaked, thus
demonstrating the advantage of the Bayesian AMP reconstruction in
terms of performance, and noise robustness. Another important point is
that the probability of a unsuccessful reconstruction is decaying
exponentially fast when the system size increases. It is also decaying
faster as $R$ decreases (see Fig.~\ref{fig:frac}).

\begin{figure*}[!ht]
\centering
\includegraphics[width=0.97\textwidth]{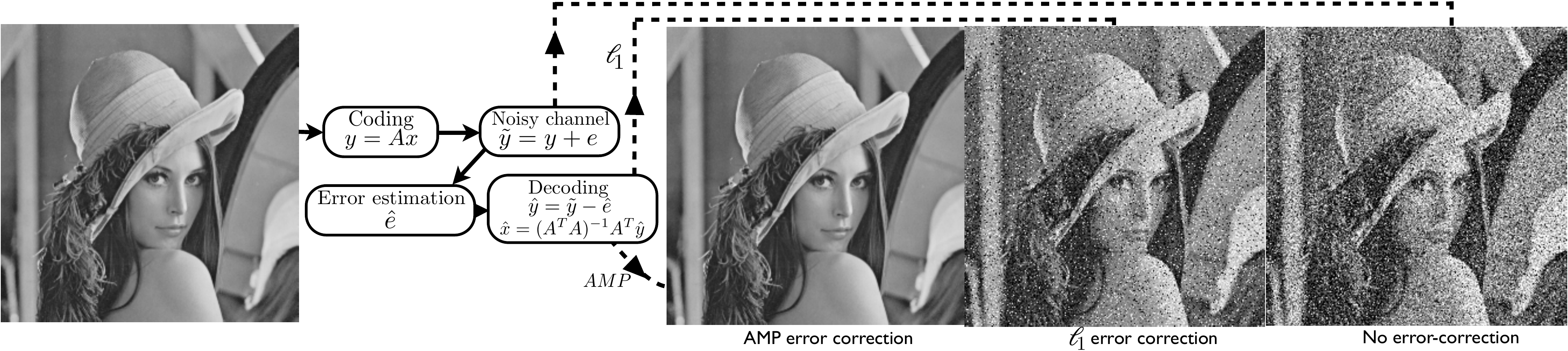}
\caption{\label{fig:Lena} Illustration of the error correcting scheme
  with the spatially coupled AMP approach and the $\ell_1$-based
  method of \cite{Candes:2008:HRE:2263476.2273354}, applied to the benchmark Lena picture. The original $256\times256$
  image is decomposed in patches of size $N=64^2$. The noisy channel
  is given by eq.~(\ref{noise}) using $\rho=0.1$, $\epsilon=10^{-6}$
  and the coding rate is $R=0.796$ (to be compared with $R_{\rm
    opt}(\epsilon=10^{-6})=0.845$) is used together with 
  spatially-coupled (seeded) matrix $F$ with parameters $L=10$, $W=3$, $J=0.2$,
  $\alpha_{\rm seed}=0.22$, $\alpha_{\rm bulk}=0.1830$.}
\end{figure*}

We also applied spatially coupled approach by choosing the matrix $F$
as described in Fig.~\ref{fig_matrix} with $L=10$, $W=3$, $J=0.2$,
$\alpha_{\rm seed}=0.22$, and varying $\alpha_{\rm bulk}$. While such
parameters are far from the limit $L\to \infty$, $W\to \infty$, $W/L
\to 0$ in which the optimal performance is guaranteed, we still obtain
a considerable improvement in the achievable coding rate, as shown
again in Fig.~\ref{fig:frac}. In Fig.~\ref{fig:Lena} we give a more
visual illustration of the performance of the spatially coupled AMP
decoder for $N=4096$. For gross error sparsity $\rho=0.1$ and small
error variance $\epsilon=10^{-6}$ we were able to perform reliable
error correction at coding rate $R=0.8$. This has to be compared with
the original approach of \cite{Candes:2008:HRE:2263476.2273354} which
only allows, even for infinite system sizes and in absence of small
noise, an asymptotic $R \approx 0.67$.

\section{Conclusion and perspectives}
We have considered an error-correcting scheme for real-valued
signals over channels that disrupt the transmitted signal by a large
error on a small fraction of elements. We combined a Bayesian AMP
reconstruction and spatially-coupled decoding matrices. We show that
this approach is robust to non-sparse small noise, we computed the phase
diagram in the limit of large signal sizes and showed numerically that
the probability of failure decreases exponentially in the signal size.

There is a number of possible improvements on the present
approach. First, the AMP approach is parallelizable, thus allowing
substantial gain in execution time. One could also use structured $F$
matrices, for instance Fourrier, Gabor or Hadamard, as in
\cite{JavanmardMontanari12}, to further decrease the execution time of the
AMP algorithm. Finally, another natural extension of the present work
is when the signal $\bx$ is itself compressible, so that a joint
source and channel coding approach can be applied. We have done
preliminary investigations on sparse signals in this direction.

\section*{Acknowledgment}
We thanks Rudiger Urbanke for useful discussions. This work has been
supported in part by the ERC under the European Union’s 7th
Framework Programme Grant Agreement 307087-SPARCS,
by the Grant DySpaN of ‘‘Triangle de la
Physique’’ and by the French Minist\`ere de la
defense/DGA.

\bibliographystyle{IEEEtran}
\bibliography{refs}

\end{document}